\documentclass{article}

\usepackage[preprint]{neurips_2026}

\usepackage[utf8]{inputenc} %
\usepackage[T1]{fontenc}    %
\usepackage{hyperref}       %
\usepackage{url}            %
\usepackage{booktabs}       %
\usepackage{amsfonts}       %
\usepackage{nicefrac}       %
\usepackage{microtype}      %
\usepackage{xcolor}         %

\usepackage{graphicx}
\usepackage{float}
\usepackage{amsmath}
\usepackage{amssymb}
\usepackage{tabularx}
\usepackage{mathtools}
\usepackage{amsthm}
\usepackage[table]{xcolor}
\usepackage[capitalize]{cleveref}
\usepackage{wrapfig}
\usepackage{enumitem}
\usepackage{booktabs}
\usepackage{titletoc}
\usepackage{tocloft}
\usepackage{xspace}

\newcommand{\pranav}[1]{{\color{black}{#1}}}
\newcommand{\pra}[1]{{\color{black}{#1}}}%
\def\ours{PINNsur\xspace}

\let\cite\citep

\title{\ours: Physics-Informed Neural Networks for PDEs on Curved Surfaces}

\author{%
  Pranav Jain \\
  University of Southern California
  \And
  Navami Kairanda \\
  Max Planck Institute for Informatics, SIC \\
  \And
  Peter Yichen Chen \\
  University of British Columbia
  \And
  Oded Stein \\
  University of Southern California \\
  Technion
}

\begin{document}

\maketitle
\newcommand{\nkc}[1]{{\color{purple}{\textbf{NK: #1}}}} %
\newcommand{\nkr}[1]{{\color{purple}{\st{#1}}}} %
\newcommand{\nke}[1]{{\color{purple}{{#1}}}} %
\newcommand{\nka}[1]{{\color{purple}{{#1}}}} %

\vspace{-30pt}
\begin{figure*}[!htb]
    \includegraphics{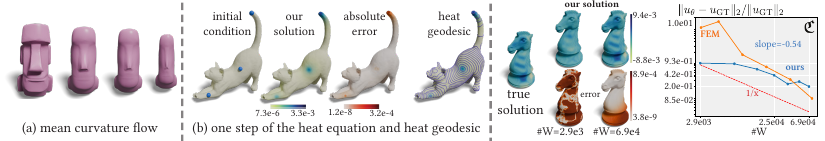}
    \vspace{-15pt}
    \caption{\ours solves PDEs on curved surfaces embedded in 3D. We empirically show that the relative $\ell_2$ error decreases as the number of tunable weights (\#$W$) of our network increases, indicating convergence of our method on arbitrary surfaces. (a) Our method simulates a mean curvature flow on a mesh. (b) We solve a step of the heat equation given initial heat sources and solve for the heat geodesic distance. (c) We then show how the predicted solution and error change as \#$W$ is increased. The true solution is the cotangent Laplacian solution on a fine mesh.}
    \label{fig:teaser}
\end{figure*}

\begin{abstract}
\pranav{Partial differential equations (PDEs) on surfaces are fundamental to scientific computing and geometry processing. A popular approach to solving PDEs on surfaces is the finite element method (FEM), where the surface is divided into discrete geometric elements (usually triangles). Recently, physics-informed neural networks (PINNs) have emerged as a continuous, mesh-free alternative that does not suffer from FEM's sensitivity to mesh quality or geometric discretization errors.
We present \ours, a simple framework for using PINNs on curved surfaces: we train a neural field to approximate the surface's normals, and then we express surface differential operators using their projection from $\mathbb{R}^3$ onto the surface. Since every orientable manifold has well-defined normals, our method is suitable for all such surfaces, regardless of curvature or topology, enabling many geometry processing applications.
Moreover, despite their empirical success in solving PDEs in flat Euclidean domains, PINNs lack convergence guarantees to the true solution of the underlying PDE, and there is limited systematic experimental evidence demonstrating such convergence. This gap restricts their adoption as reliable solvers compared to established methods like FEM, where convergence to the true solution is well understood and theoretically grounded. 
These surface PDEs are particularly challenging to solve convergently, as one must not only deal with the convergence of the function approximation, but also with the convergence of the geometric approximation of the surface itself.
In this work, we empirically investigate the convergence behavior of PINNs for solving surface PDEs by introducing a simple empirical convergence test.}
\end{abstract}

\section{Introduction}
\vspace{-9pt}
Partial Differential Equations (PDEs) on curved two-dimensional surfaces \(\Omega \subseteq \mathbb{R}^3\) are at the core of many 3D geometry processing applications such as surface editing, animation, physical simulation, and computer-aided manufacturing.
Consequently, accurate, efficient, and easy-to-implement solvers for surface PDEs have become a central component of modern geometry processing pipelines.
Curved surfaces pose additional challenges to PDE solving compared to flat two-dimensional regions \(\subseteq \mathbb{R}^2\) or \(\subseteq \mathbb{R}^3\):
One must not only discretize the PDE, but also the surface itself.
An extremely common discretization of surfaces has been the triangle mesh, which works well together with the Finite Element Method (FEM) for discretizing PDEs, especially linear (Lagrangian) FEM, which approximates solutions using piecewise linear functions defined on triangular elements.
\pranav{Linear FEM has many advantages, the main one being that it can solve the Poisson equation in a convergent way given certain conditions \cite{Dziuk1988,Wardetzky2007DiscreteDifferentialOperators}.
It also has a variety of drawbacks, such as needing meshes with well-behaved triangles and careful refinement.}

To address FEM's drawbacks, Physically Informed Neural Networks (PINNs) have become popular \cite{karniadakis2021physics,hao2022physics}, \pranav{where the function is approximated by a differentiable neural network, and the PDE is solved using \textit{autodifferentiation}.}
Neural network training methods are then used to find the weights that solve the PDE.
PINNs address some of FEM's shortcomings, such as the dependence on a mesh with high-quality triangles, but are difficult to extend to curved surfaces: The coordinate networks underlying PINNs generally only work on \(d\)-dimensional flat subsets of \(\mathbb{R}^d\) (i.e., areas in \(\mathbb{R}^2\) and volumes in \(\mathbb{R}^3\)).
\pranav{Recent works \cite{williamson2025,welschinger2025learning} have begun to extend PINNs to curved surfaces and work towards integrating PINNs in mainstream geometry processing applications.}
One line of work formulates PINNs in \(\mathbb{R}^3\), but explicitly evaluates a surface differential operator (such as the Laplace-Beltrami operator \(\Delta\)) using standard formulas from differential geometry.
The work of \citet{williamson2025}, in particular, does this by parameterizing the surface \(\Omega\) using a sphere as a parameter domain, and using that parameterization to construct a Laplace-Beltrami operator.
There is no explicit discretization of the PDE or the surface geometry into elements; instead, the effective discretization is implicitly controlled by the number of network weights. \pranav{Thus, unlike FEMs, which provide convergence guarantees to the true solution under consistent discretization (typically as a function of the number of triangular elements $n$), neural solvers lack analogous guarantees, and systematic studies remain limited.}

\pranav{Building on the ideas of \citet{williamson2025}, we introduce a new framework for using PINNs on curved surfaces by completely eliminating the parameter domain, simplifying the network, and extending it to arbitrary manifold topologies.}
Our key observation is that common surface PDE operators (e.g., Poisson and Helmholtz) do not require a parametric mapping of the surface; normals alone suffice to construct surface differential operators, since one can define the operators in Euclidean space and project out normal components to obtain the surface operators.
\pranav{Our method supports both Dirichlet and Neumann boundary conditions and performs well on many challenging geometries.}
\pranav{We then study convergence behavior to the true solution for the proposed surface PDE solver.
We introduce a notion for the convergence of PINNs by relating the error to the number of degrees of freedom in the PINN.
Our convergence criterion parallels classical convergence criteria in use for FEM, where the error is bounded by an expression containing the number of elements or the element size (both proxies for the total number of degrees of freedom). Analogously, for PINNs, the number of trainable parameters can be interpreted as an effective measure of degrees of freedom.
We study both classical PINNs as well as our new PINN for curved surfaces using this criterion, and establish their convergence.
}

Our contributions are :
\begin{enumerate}[noitemsep,topsep=0pt] %
    \item A simple surface PINN framework called \ours for solving PDEs using a normals-only neural field for surface operators.
    \pranav{\item A new empirical convergence criterion for PINNs.}
    \pranav{\item An empirical systematic convergence study of PINNs on test functions across varied PDEs, geometries, and boundary conditions.}
\end{enumerate}
Our complete code can be accessed here: \href{https://github.com/Pranav-Jain/NN_convergence.git}{code link}.

\section{Related Work}
\label{sec:relatedwork}
\paragraph{Classical Methods for Surface PDEs.}
There are many classical methods for solving surface PDEs; we highlight only a few important examples that are used in the geometry processing of curved surfaces.
\citet{Dziuk1988} proved an early convergence result for the linear FEM on curved surfaces approximated by triangle meshes, which was later extended by \citet{Wardetzky2007DiscreteDifferentialOperators}.
These results generalize known flat FEM theorems \cite{brennerscott}.
As a rule of thumb, such FEM methods have \(L^2\) convergence on the order of \(\mathcal{O}(n)\), where \(n\) is the number of degrees of freedom.
The reliability of linear FEM has led to its popularity in many geometry processing applications such as the computation of minimal surfaces \cite{pinkallpolthier}, surface fairing \cite{Desbrun1999}, mesh editing \cite{Sorkine2004}, animation \cite{Baran2007}, and many more \cite{polygonmeshprocessing}.
Beyond linear FEM, other popular discretization methods include higher-order FEM \cite{Ferguson2023}, discrete exterior calculus \cite{hiranidec}, and Monte Carlo methods (which have only recently been extended to surfaces) \cite{Sugimoto2024} (which converge on the order of $\mathcal{O}(\sqrt{n})$, where \(n\) is the number of degrees of freedom).

\paragraph{Physics Informed Neural Networks.}
PINNs are neural fields that model solutions to differential equations by posing them as optimization problems supervised by physical laws \citep{karniadakis2021physics, hao2022physics, raissi2019physics, Kharazmi2021}. 
They offer a flexible alternative to classical discretization-based solvers, especially when data or meshing is challenging. 
PINNs have been applied across a range of problems, including fluid dynamics \citep{cai2021physics,jain2024neural}, volumetric elastodynamics \cite{rao2021physics}, Eikonal equations \citep{smith2021eikonet}, and cloth simulation \cite{kairanda2023neuralclothsim, kairanda2026dinf}. 
Beyond direct neural solvers, there is growing interest in \emph{neural operators} that learn solution maps for parametric PDEs \cite{yu2018deep, welschinger2025learning}.
Early PINN works with coordinate-based MLPs struggled with high-frequency signals and derivatives; Siren \citep{sitzmann2020implicit} mitigates this via periodic activations and has enabled broader applications \citep{chen2023implicit,zehnder2021ntopo,kairanda2023neuralclothsim}. 
However, Siren and most PINNs are formulated on Euclidean grids rather than surfaces. A concurrent work by \cite{welschinger2025learning} solves PDEs on surfaces via a \emph{neural operator}, but still uses Euclidean finite differences. In contrast, our solver models surface operators directly and solves the PDE on the continuous surface field.

Previous work on PINN convergence includes
\citet{shin2020convergence}, which prove that, under certain conditions, PINNs on flat domains in \(\mathbb{R}^n\) converge to their solutions, and provide experimental evidence that increasing the amount of training data decreases error.
\citet{tang2021physics} investigate the convergence behavior of PINNs on curved surfaces as the number of training samples increases, and they do so for different network depths.
\citet{nam2024solving} introduces a novel neural Monte Carlo method and investigates its convergence behavior as the network is trained over time.
\pranav{Prior work focuses on training convergence of PINNs, whereas we study convergence with respect to degrees of freedom, where the number of tunable weights plays a role analogous to FEM elements.}

\paragraph{Neural Geometry Processing.}
Many recent methods encode geometric shapes (e.g., surfaces) as neural networks, often as neural fields, enabling geometry processing tasks \cite{yang2021geometry,bednarik2020shape, mehta2022level,Gao2024,Edavamadathil2024,Fargion2025,chetan2025accurate}. 
A key advantage is that differential geometry quantities like normals and curvatures can be computed via \emph{autodiff}. \citet{morreale2021neural} learn explicit surface maps for distortion-minimizing correspondences, and \citet{morreale2022neuralconvolutionalsurfaces} disentangles fine detail from global structure, while \citet{Aigerman2022Neural} learns piecewise-linear mesh mappings.
\citet{novello2022exploring,Novello2023neural} explore differentiable geometry of level-set surfaces and how to evolve implicit level-set functions, while \citet{yang2021geometry} demonstrates surface smoothing and sharpening on implicits.
Recently, \citet{chang2025shape, williamson2025} solve the Laplace--Beltrami operator to perform spectral analysis. The closest work to ours is \citet{williamson2025}; the key difference is that we avoid a parameterized surface map, since it is not part of the surface PDE operator. We instead use a normal field, which improves accuracy and convergence.

\section{Method}
\vspace{-10pt}
Our setting is a (curved) two-dimensional manifold surface \(\Omega \subseteq \mathbb{R}^3\).
We solve PDEs of the form
\begin{align}
    \mathcal{F}[u(\mathbf x)] = 0,\quad & \mathbf{x}\in \Omega,\\
    u(\mathbf x) = g(\mathbf x),\quad &\mathbf x\in \partial\Omega_D,\\
    \frac{\partial u}{\partial \mathbf \nu}(\mathbf x) = h(\mathbf x), \quad &\mathbf x\in \partial \Omega_N,
\end{align}
where $\mathcal{F}$ is a second-order elliptic differential operator, $\partial\Omega_D$ is the boundary curve with Dirichlet condition and $\partial\Omega_N$ with Neumann condition with $\mathbf \nu$ as the boundary normal.
The no-boundary case, \(\partial\Omega_D = \partial\Omega_N = \varnothing\), occurs for some curved surfaces.
In this work, we focus on the Poisson and Helmholtz equations because of their wide applications \cite{Kazhdan2013, pinkallpolthier}.

\subsection{Poisson Equation}
The (Euclidean) Laplacian $\Delta_{\mathbb R^d}$ of a scalar field $u: \mathbb{R}^d \to \mathbb{R}$ ($d=2,3$) is defined as the divergence of the gradient of the field.
The Laplace-Beltrami operator, or surface Laplacian, is the generalization of the Laplacian $\Delta_{\mathbb R^d} u$ to curved domains $\Omega$,
\(
    \Delta_\Omega u =\nabla_\Omega\cdot\nabla_\Omega u.
\)
The surface gradient is the orthographic projection of the Euclidean gradient to the local tangent plane
\(
    \nabla_\Omega u = \nabla_{\mathbb R^d} u - (\nabla_{\mathbb R^d} u\cdot \mathbf n)\mathbf n.
\)
Similarly, for a vector field $\mathbf{F}: \mathbb R^d \to \mathbb R^d$, the surface divergence is defined as 
\begin{equation}
    \nabla_\Omega \cdot \mathbf{F} = \nabla_{\mathbb R^d} \cdot \mathbf{F} - \mathbf{n^TJ(F)n}, \text{ where }
    \mathbf{J(F)} = \left[ \frac{\partial \mathbf F}{\partial x_1}\quad \frac{\partial \mathbf F}{\partial x_2} \cdots \frac{\partial \mathbf F}{\partial x_d} \right]
\label{eq:surface_divergence}
\end{equation}
is the Jacobian of $\mathbf F$ \cite{hubbard:hal-01297648}.%

The Poisson equation on a curved domain $\Omega \subset \mathbb R^d$ for a scalar function $u: \Omega \to \mathbb R$ is given by
\(
    \Delta_\Omega u = f ,
\)
where $f: \Omega \to \mathbb R$ is the source term.
It is relatively straightforward to compute the Euclidean Laplacian of a coordinate network by applying \textit{autodifferentiation} to a smooth network \(u: \mathbb{R}^d \rightarrow \mathbb{R}\):
\begin{equation}
    \Delta_{\mathbb R^d} u = \frac{\partial^2 u}{\partial x_1^2} + \frac{\partial^2 u}{\partial x_2^2} +\cdots +\frac{\partial^2 u}{\partial x_d^2}
    \;.
\end{equation}
\pranav{This simple formula does not apply to the Laplace-Beltrami operator, where we can not simply employ component-wise differentiation.}
\pranav{The approach of \citet{williamson2025} is to represent the surface with a parametric neural map \(\varphi_\theta : \mathbb{S}^2 \rightarrow \Omega\).
\(\varphi_\theta\) is obtained by first computing a discrete parametrization of the surface triangle mesh from a reference sphere, and then training a neural network to approximate that discrete parametrization.}
While this allows computing normals via the Jacobian of $\varphi_\theta$, it does not handle arbitrary topologies (the topology of the parameter domain must match the topology of \(\Omega\)).
\pranav{In contrast, our method does not require a parametrization at all.}
We also represent \(u\) as a coordinate network, but instead of a parametrization we directly work with the normal field that is either known exactly, or that we represent with another network that can be trained quickly and used without any autodifferentiation.
The resulting formula for the Laplace-Beltrami operator of a coordinate network is:
\begin{align}
    \label{eq:howlaplacebeltramicomputed}
    \Delta_\Omega u &= \nabla_{\mathbb R^d} \cdot (\nabla_{\Omega}u) - \mathbf{n^{\top}J}(\nabla_{\Omega}u)\mathbf{n}
    \;.
\end{align}

The solution of the actual PDE works as follows.
Given a function $f$ and a surface $\Omega$,
we train a PINN to learn the function $u$ by defining the loss function $\mathcal{L}_{\text{poisson}}$ as 
\begin{equation}
\mathcal{L}_{\text{poisson}} = \| \Delta_\Omega u_\theta - f \|_2^2.
    \label{eq:loss_poisson}
\end{equation}
The Laplace-Beltrami operator ($\Delta_\Omega$) is computed using \eqref{eq:howlaplacebeltramicomputed}. Please see \cref{fig:overview} for an overview.

\subsection{Helmholtz Equation}
The Helmholtz equation on a curved domain $\Omega \subset \mathbb R^d$ for a scalar function $u: \Omega \to \mathbb R$ is defined as $\Delta_\Omega u + k^2u = f$,
where $k$ is the wavenumber and $f: \Omega \to \mathbb R$ is the source term.

We use the same framework as for the Poisson equation, but define the loss function $\mathcal L_{\text{helmholtz}}$ as
\begin{equation}
    \mathcal L_{\text{helmholtz}} = \| \Delta_\Omega u_\theta + k^2 u_\theta - f \|_2^2.
    \label{eq:loss_helmholtz}
\end{equation}

\subsection{Boundary Conditions}
We employ so-called soft boundary constraints \cite{raissi2019physics,chen2023implicit}, where boundary conditions are enforced by adding penalty terms to the loss function during training.
\begin{equation}
\mathcal{L}_{\text{soft}} =
\mathcal{L}_{\text{PDE}} +
\lambda_{\text{BC}} \mathcal{L}_{\text{BC}},
\end{equation}
where $\lambda_{\text{BC}}$ is a tunable parameter.
We choose $\lambda_{\text{BC}} = 100$ to give a high weight to the boundary conditions. \cref{fig:error_vis} shows that the solution error at the boundary is less dependent on the parameter $\lambda_{\text{BC}}$, but the error at the boundary decreases as the size of the network increases.

Different boundary conditions necessitate different loss functions \(\mathcal{L}_{\text{BC}}\).

\paragraph{Dirichlet Boundary.} To solve a PDE with Dirichlet boundary condition $u(\mathbf{x}) = g(\mathbf{x}) , \mathbf{x}\in \partial\Omega_D$, \pranav{we add the following term to the loss function.}
\begin{align}
    \mathcal{L}_{\text{BC}} = \| u_\theta - g \|_2^2.
    \label{eq:dirichlet_term}
\end{align}

The loss in \eqref{eq:dirichlet_term} is minimized exactly when the Dirichlet boundary conditions hold.
Since a valid solution to a PDE with Dirichlet boundary conditions exist, there must be a global minimum that both minimizes the main PDE loss (\eqref{eq:loss_poisson}, \eqref{eq:loss_helmholtz}), as well as the additional boundary loss \eqref{eq:dirichlet_term}.

\paragraph{Neumann Boundary.} To enforce Neumann boundary conditions, one needs to compute $\frac{\partial u}{\partial \mathbf \nu}$.
Since we enforce Neumann conditions with a soft loss term involving $\frac{\partial u}{\partial \mathbf \nu}$, this quantity needs to be differentiable, which we achieve by training another coordinate MLP on the boundary normals \(\nu\) by computing the cross product between the boundary edge and the surface normal at the boundary.
\pranav{Then, we use this MLP to compute the following term at $\partial \Omega_N$}
\begin{equation}
    \mathcal{L}_{\text{BC}} = \left\| \nabla_\Omega u_\theta\cdot \mathbf \nu_{\tilde \theta} - h \right\|_2^2.
    \label{eq:neumann_term}
\end{equation}

\paragraph{Compatibility.}
When solving for the Poisson or Helmholtz equations, the compatibility condition is required to ensure the existence of a solution \cite{Evans2010}. Existence and uniqueness are guaranteed for both Poisson and Helmholtz when pure Dirichlet boundary conditions are imposed (as long as the boundary condition is compatible with the interior function values). For the Poisson equation $\Delta_{\Omega} u = f$ and the Helmholtz equation $\Delta_{\Omega} u + k^2u= f$ with a Neumann boundary, the solution exists and is unique up to a constant if
\begin{equation}
    \int_\Omega f +  \int_{\partial \Omega} \frac{\partial u}{\partial \mathbf \nu} = 0.
    \label{eq:compatibility}
\end{equation}
For surfaces without a boundary, we simply have to ensure $\int_\Omega f = 0$. For all our experiments on surfaces with Neumann or no boundary, we choose the source terms \(f\) that ensure homogeneous Neumann boundary  \pranav{($h(\mathbf x)=0$)} and $\int_{\Omega} f = 0$. 

\subsection{Network Design \& Training}
\label{sec:network_design}

\paragraph{Normal Networks.}
To compute the Laplace-Beltrami operator \eqref{eq:howlaplacebeltramicomputed}, we need the surface normals $\mathbf n$ at each point on the surface $\Omega$ and boundary normals $\nu$ for Neumann boundary conditions.
We use a \textit{Siren} network \cite{sitzmann2020implicit} which is a multilayer perceptron with \(sine\) activation functions to train both the surface normal field and the boundary normal field. We chose \textit{Siren} because of its ability to approximate high-frequency functions (\cref{fig:activation_fn}).

\pranav{Our surface normal network is $\mathbf{n}_{\hat\theta}: \mathbb{R}^d \to \mathbb R^d$, mapping the Euclidean coordinates $\mathbf x\in \Omega$ to the normal $\mathbf n$ at $\mathbf x$ (even though the coordinate network takes points in \(\mathbb{R}^3\) as input, we only ever evaluate it on \(\Omega\)).
The last layer of the network normalizes the output to ensure unit normals. Therefore, while training, we focus only on the direction of the normals at each point on $\Omega$. We define the loss function}
\pranav{
\begin{equation}
    \mathcal{L}_{\text{normals}} = \sum_{i=1}^N (1-(\mathbf{n}_{\hat\theta_i} \cdot \mathbf{n}_{\text{GT}_i}))^2
    \;,
    \label{eq:loss_normals}
\end{equation}

where $N$ is the number of uniformly sampled points on the surface $\Omega$.
In \cref{fig:normal_network_params}, we illustrate how the PDE approximation error changes as the size of the normal network increases. The error initially decreases and then plateaus, indicating that once the network is sufficiently expressive to approximate the normal field accurately, further increases in its size no longer impact the error. Empirically, we observe that a network with width $64$ and depth $5$ is adequate for learning the normal field.
We train the network for $\approx 1e7$ iterations with a learning rate of $1e^{-4}$. The normal network training converges quickly for simple domains but takes more time for complex geometries.}

Our boundary normal network \pranav{$\nu_{\tilde\theta}: \mathbb{R}^d \to \mathbb R^d$} takes the Cartesian coordinates $\mathbf x \in \partial\Omega_N$ and outputs the boundary normal \(\nu\) at $\mathbf x$ (even though the coordinate network takes points in \pranav{\(\mathbb{R}^d\)} as input, we only ever evaluate it on \(\partial\Omega_N\)).
The training \pranav{hyperparameters} for $\nu_{\tilde\theta}$ are the same as for $\mathbf n_{\hat \theta}$.
We use the same loss function
\eqref{eq:loss_normals} where $\nu_{\text{GT}}$ is computed using the cross product between the edge and the surface normal.

\begin{wrapfigure}[16]{r}{0.45\textwidth}
    \vspace{-20pt}
    \begin{center}
        \centering
        \includegraphics[width=0.85\linewidth]{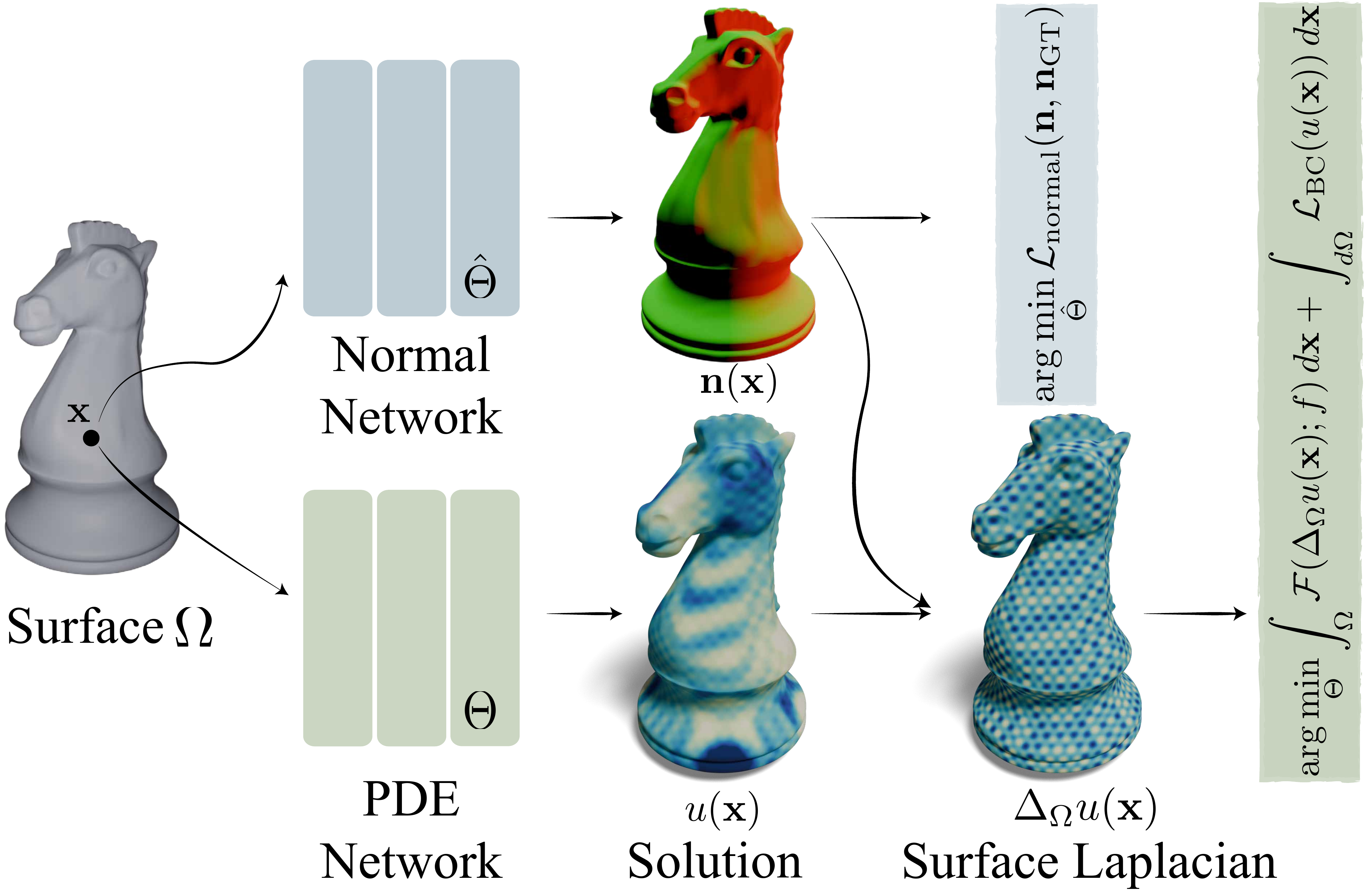}
        \caption{\pranav{\textbf{Overview}. To solve PDEs on a curved surface, we first train a normal network to approximate the surface normals. The PDE network then leverages these normals to compute the surface Laplacian $\Delta_\Omega u$, with boundary conditions added as soft constraints.}}
        \label{fig:overview}
    \end{center}    
\end{wrapfigure}

\paragraph{PDE Network $u_\theta$.} Once we have the surface normals (and boundary normals) we compute the Laplace-Beltrami operator using \eqref{eq:howlaplacebeltramicomputed}. Note that since the normals are computed using differentiable neural networks $\mathbf n_{\tilde \theta}$, $\nu_{\tilde \theta}$, we can use \emph{autodiff} to train for $u_\theta$. 
We train our network $u_\theta$ to solve for the Poisson and the Helmholtz equation using \eqref{eq:loss_poisson} and \eqref{eq:loss_helmholtz} respectively. The boundary loss terms \eqref{eq:dirichlet_term} and \eqref{eq:neumann_term} are added as soft constraints. \cref{fig:overview} shows an overview of our method. 

For all experiments, we choose a starting learning rate of $1e^{-3}$ with the \textit{ReduceLROnPlateau} scheduler. We uniformly sample points on the domain and train until $u_\theta$ converges and saturates completely. Empirically, we noticed that the training usually converges around $50k$ iterations (\cref{fig:loss_plots}). We train all the networks for $150k$ iterations to ensure the network is completely saturated.

To show convergence, we train our PDE network $u_\theta$ with depth $3$ and vary the width of the network from $30$ to $150$. \pranav{In \cref{fig:width_depth}, we show that the choice of varying depth versus width doesn't change the final error. Therefore, without loss of generality, we choose to vary the width in all our experiments.} We consider the degrees of freedom to be the total number of tunable weights+biases in the network (\#$W$), which is analogous to the number of vertices in a mesh for FEM. Please see \cref{sec:convergence} for further details.

\begin{figure*}
    \centering
    \includegraphics{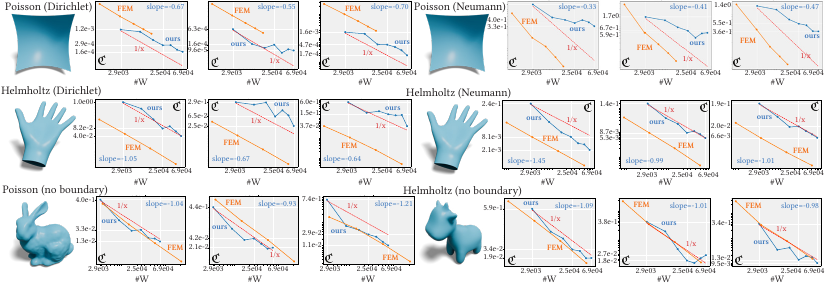}
    \vspace{-20pt}
    \caption{Convergence plots for the Poisson and Helmholtz equations with different boundary conditions for three different functions defined on each surface. For all surfaces and boundary conditions, we show that our method is convergent.}
    \label{fig:convergence_plots}
\end{figure*}

\section{Convergence}
\label{sec:convergence}

To quantify the accuracy of a PDE solver, we need a notion of convergence. Convergence describes whether the estimate approaches the true solution with respect to a chosen norm. Here we use the relative $\ell_2$ norm and define error as
\begin{equation}
    e :=\| u_\theta - u_\text{GT}\|_2 / \| u_\text{GT} \|_2\; ,
    \label{eq:error}
\end{equation}
where $u_\theta$ and $u_\text{GT}$ are our optimized solution and the ground-truth, evaluated respectively on $\mathbf{x} \in \Omega$.

In classical discretization-based solvers such as
finite-element schemes, convergence is often expressed in relation to refining the number of elements (\(\sim\) degrees of freedom) \(n\) of the discretization (or a quantity similar to the element size) \cite{brennerscott}.
Convergence is observed if the error decreases as more elements are added to the discretization.
In our method (and more broadly in neural solvers that model continuous fields), there is no discretization in the classical sense (and therefore no direct refinement analog).
Instead, the weights of the neural field induce an implicit discretization by limiting the model's expressive capacity to represent the continuous solution.
We therefore treat the total number of tunable neural network weights (\#$W$) as degrees of freedom and study convergence under the expectation that the error $e\to 0$ as \#$W \to \infty$. 
We compute \#$W$ as the number of parameters of the \emph{Siren} network that models the PDE solution $u_\theta$. Since \emph{Siren} is a coordinate-based MLP, \#$W$ is determined by the depth and width of the MLP. We exclude the normal network $\mathbf{n}_{\hat\theta}$ from this count, since it is pretrained and unaffected by the PDE. For $u_\text{GT}$, we use analytical solutions when available; otherwise, we use a high-resolution FEM solution as GT (cotangent Laplacian \cite{Meyer2002DiscreteDO} on the surface $\Omega$). 

With the above setup, we define the following criterion: if the error $e$ in \eqref{eq:error} systematically decreases as the number of tunable weights increases, the method converges.
To compute this, we fit a line via least-squares linear regression of $\log e$ on $\log($\#$W)$ and obtain the slope $m$ and Pearson correlation coefficient $r$. An experiment is deemed to have converged (indicated by $\mathfrak{C}$ in figures) if $m < -0.3$ and \pranav{$r < -0.5$}, indicating a sufficiently steep decay and a reasonably linear trend.  \pranav{We emphasize that the chosen convergence heuristic is purely empirical, chosen to promote a stable and monotonic decrease in approximation error as the degrees of freedom increase, while mitigating oscillations arising from noise. The use of a lower convergence rate is consistent with previous works in certain areas of numerical analysis, such as \citet{Li2018, Apel2018}.}
\vspace{-4pt}

\begin{figure*}
    \centering
    \includegraphics{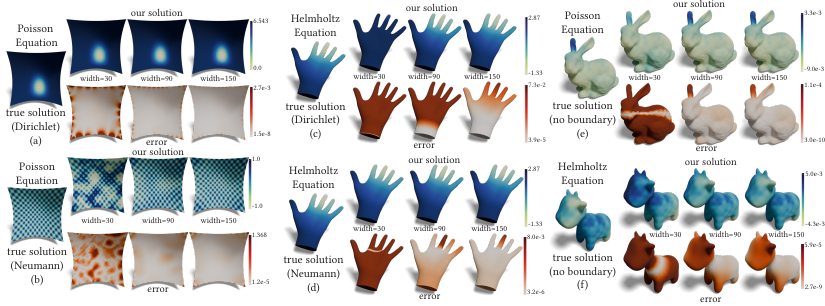}
    \vspace{-20pt}
    \caption{\pranav{Figure showing the solution and error as network width increases for Poisson and Helmholtz equations under various boundary conditions. We observe that error decreases with increasing width.}}
    \vspace{-10pt}
    \label{fig:error_vis}
\end{figure*}

\section{Results}
\vspace{-5pt}
\subsection{Solving PDEs}

We solve the Poisson equation on a wide variety of domains:
flat surfaces \(\subseteq \mathbb{R}^2\), where our method simplifies to a standard PINN (\cref{fig:comparison_baselines}), curved surfaces with boundary where exact normals and boundary normals are known (\cref{fig:error_vis}(a, b)), and arbitrary manifold surfaces given as triangle meshes, where we train a neural network to have a smooth normal field that approximates the piecewise constant mesh normals (\cref{fig:error_vis}(e)).
Beyond the Poisson equation, our method also solves the Helmholtz (\cref{fig:error_vis}(c, d, f)) and heat equations (\cref{fig:teaser}(b)).
\pranav{In \cref{fig:intricate}, we demonstrate our ability to solve the Helmholtz equation on a challenging intricate curved mesh with fine geometric detail.
The details of the experimental setup are provided in \cref{sec:experiment_setup}.}

\subsection{Convergence Experiments}

\pra{Using our convergence criterion established in \cref{sec:convergence}, we evaluate and observe the convergence of our method for both Poisson and Helmholtz equations on a wide variety of surfaces with different boundary conditions using different source terms (see \cref{fig:convergence_plots}). 
We compare convergence rates with \citet{Kharazmi2021, yu2018deep, kairanda2026dinf} and FEM on a flat 2D domain, matching FEM convergence rate (\cref{fig:comparison_baselines}). On curved surfaces, we compare against FEM and a parameterized PINN \citep{williamson2025} (see \cref{sec:baselines} for more details).
The runtimes for each convergence experiment in \cref{fig:error_vis} can be found in \cref{sec:timings} of the supplemental.}

\subsection{Ablation Study}
\label{sec:ablation_study}
To justify our parameter choices, we perform ablations.
In all our experiments, we vary the network width while keeping the depth constant. \cref{fig:width_depth} shows that the choice of altering depth instead of width gives similar results. In \cref{fig:activation_fn}, we show that the \emph{Siren} network is much better at learning high-frequency functions compared to using \emph{sigmoid} or \emph{tanh} activation functions. 
\pranav{In \cref{fig:normal_network_params}, we research the dependence of the error on the normal network size.
The error initially decreases and then remains constant, indicating that once the normal network is deep enough, only the main network weights matter.}

\begin{figure}
    \centering
    \includegraphics{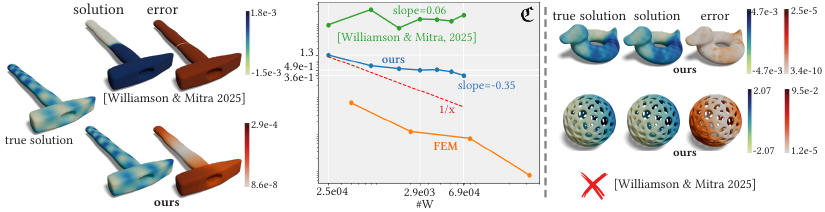}
    \vspace{-20pt}
    \caption{\pranav{\textbf{Comparison with \citet{williamson2025}}. By directly training on surface normals, our method achieves a lower PDE solution error. Moreover, our approach applies to arbitrary topologies, whereas \citet{williamson2025} is limited to surfaces homeomorphic to a sphere.}}
\label{fig:williamson_comparison}
\vspace{-10pt}
\end{figure}

\subsection{Comparison with Baselines}
\label{sec:baselines}

\begin{figure}
    \centering

    \begin{minipage}{0.5\textwidth}
        \centering
        \includegraphics[width=\linewidth]{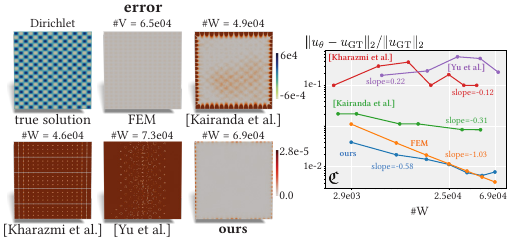}
    \end{minipage}\hfill
    \begin{minipage}{0.5\textwidth}
        \centering
        \scriptsize
        \begin{tabular}{c|ccc}
          \rowcolor{blue!20}
          \underline{Method} &
          \multicolumn{3}{c}{\underline{Degrees of Freedom}} \\
          \rowcolor{blue!20}
          & 2.91e3 & 2.49e4 & 5.16e4 \\ \hline\hline
          FEM & $1.11\mathrm{e}{-1}$ & $7.78\mathrm{e}{-3}$ & $\mathbf{5.55\mathrm{e}{-3}}$ \\
          \cite{kairanda2026dinf} & $1.98\mathrm{e}{-1}$ & $9.67\mathrm{e}{-2}$ & $8.06\mathrm{e}{-2}$ \\
          \cite{Kharazmi2021} & $1.00\mathrm{e}{0}$ & $1.83\mathrm{e}{0}$ & $9.99\mathrm{e}{-1}$ \\
          \cite{yu2018deep} & $1.76\mathrm{e}{0}$ & $5.04\mathrm{e}{0}$ & $4.57\mathrm{e}{0}$ \\
          \textbf{ours} & $\mathbf{3.97\mathrm{e}{-2}}$ & $\mathbf{1.13\mathrm{e}{-2}}$ & $6.04\mathrm{e}{-3}$ \\
          \bottomrule
        \end{tabular}
    \end{minipage}
    \vspace{-5pt}
    \caption{\textbf{Poisson solve on a Euclidean domain}. A comparison with FEM, \citet{kairanda2026dinf}, \citet{Kharazmi2021}, \citet{yu2018deep} for similar degrees of freedom shows that the relative $\ell_2$ error and convergence for ours match more closely with FEM. %
    }
    \vspace{-5pt}
    \label{fig:comparison_baselines}
\end{figure}

Our PINN exhibits a significantly reduced sensitivity to mesh quality, unlike FEM (\cref{fig:bad_mesh}).
\pra{Classical PINNs such as \citet{Kharazmi2021, yu2018deep} cannot directly solve surface PDEs, so we compare FEM, \citet{kairanda2026dinf,Kharazmi2021,yu2018deep}, and our method on \(\mathbb{R}^2\).
Our approach exhibits convergence behavior similar to FEM for comparable degrees of freedom (\cref{fig:comparison_baselines}), while \citet{Kharazmi2021,yu2018deep} do not converge under the network refinement strategy used here.
We attribute this difference to the use of Siren, which better captures the high-frequency solution $u$ compared to the \emph{tanh} of \citet{Kharazmi2021} and $ \max(0, x^2)$ of \citet{yu2018deep}.
Although \citet{kairanda2026dinf} demonstrates convergence with increasing resolution, improvements eventually saturate due to interpolation assumptions.

On surfaces $\Omega \subset \mathbb{R}^3$, we compare against a parameterization-based approach similar to \citet{williamson2025} (\cref{fig:williamson_comparison}).
We speculate that their weaker convergence stems from challenges in learning parameterizations and differentiation errors, both avoided by our method. Note that we support arbitrary manifold topologies, unlike \citet{williamson2025}. We also compare with \citet{Sugimoto2024}, a Monte Carlo method limited to Dirichlet problems with simple source terms, while our approach supports both Dirichlet and Neumann boundaries. (\cref{fig:comparison_sugimoto})}

\subsection{Geometry Processing Applications}

Since our method works reliably on a wide variety of curved surfaces, it can be used directly for a wide range of geometry processing applications where FEM would traditionally be used.

\begin{wrapfigure}[17]{r}{0.4\textwidth}
    \begin{center}
        \centering
        \vspace{-20pt}\includegraphics[width=1.0\linewidth]{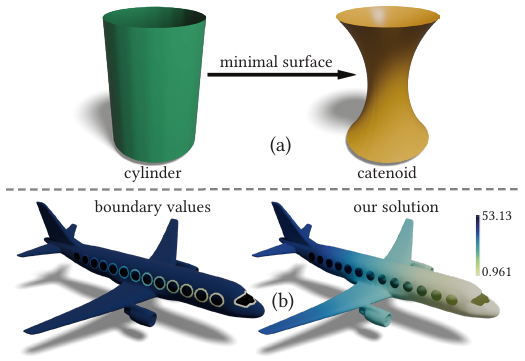}
        \vspace{-20pt}
        \caption{\pranav{(a) Result of the minimal surface computed for the cylinder. (b) Result of harmonic interpolation in the interior of the plane given Dirichlet boundary values on the windows of the plane.}}
        \label{fig:gp_combined}
    \end{center}    
\end{wrapfigure}

\paragraph{Minimal Surfaces.} 
A surface $\Omega \subset \mathbb{R}^3$ is minimal if its mean curvature is zero everywhere. The mean curvature relates to the Laplace–Beltrami operator via $\Delta_\Omega = -2H\mathbf{n}$, where $H$ is the mean curvature and $\mathbf{n}$ the surface normal.
To compute a minimal surface from a cylinder, we solve three Laplace equations: $\Delta_\Omega u = 0$ with Dirichlet boundary conditions for each coordinate $(x,y,z)$. The resulting solutions define the new vertex positions. This recovers the cylinder’s minimal surface, the catenoid (\cref{fig:gp_combined}(a)).
\paragraph{Harmonic Interpolation.} In \cref{fig:gp_combined}(b), we show that \ours can be used to compute the harmonic interpolation of a surface given the Dirichlet boundary condition by solving the Laplace equation
\begin{equation}
    \Delta_\Omega u = 0, \quad u(\mathbf x) = g(\mathbf x), \mathbf x\in \partial\Omega_D \;.
\end{equation}
Here, our method's ability to handle all domain topologies particularly shines, since interpolation examples can have many different disconnected boundaries.

\paragraph{Heat Equation.} Laplacians can also be used to solve the heat equation. The implicit Euler finite difference scheme in time for the heat equation $\frac{\partial u}{\partial t} = \Delta_\Omega u,$ results in a screened Poisson equation
\begin{equation}
    \Delta_\Omega u^{t+1} - h^{-1}u^{t+1} = -h^{-1}u^t \;,
    \label{eq:screened_poisson}
\end{equation}
where $h$ is the time-step size. \ours solves the heat equation on intricate curved surfaces by training for each time step. One-step solution of the heat equation given initial heat sources is shown in \cref{fig:teaser}(b). \pranav{We can also compute heat-based geodesic distances on a surface using the method of \citet{heatgeodesic} as shown in \cref{fig:teaser}(b).} 

\paragraph{Mean Curvature Flow.} Using the heat equation, we can also simulate a mean curvature flow (\cref{fig:teaser}(a)) by solving \eqref{eq:screened_poisson} three times for $u^t = {x, y, z}$ where ${x, y, z}$ are the mesh vertex coordinates. The solution $u^{t+1}$ forms the new vertex coordinates.

\section{Limitations}
\vspace{-5pt}
\label{sec:limitations}
While we show convergence of our method on a range of surfaces and boundary conditions, our method fails if the surface has a discontinuous normal field. For example, as shown in \cref{fig:limitations}, the boundary normals are discontinuous near the corners, and hence the model fails to converge when solving a Poisson equation with Neumann boundary.
Our method also fails to converge if the solution $u$ has a high frequency or if the range of the function is too high.
\pranav{Note that our method can sometimes have a higher error than FEM
and incur greater computational cost due to training, similar to other PINNs, but does not require high-quality meshes like FEM.
}
\vspace{-5pt}

\section{Conclusion \& Discussion}
\vspace{-5pt}
\pranav{We introduced \ours for solving second-order PDEs like the Poisson or Helmholtz equation on curved surfaces.
\ours works on manifold surfaces of arbitrary topology (unlike prior PINNs, which either do not support surfaces at all or are limited to genus-0 surfaces), and offers better convergence behavior than state-of-the-art PINNs.
We also introduce an empirical convergence criterion for PINNs, generalizing the principle used for determining the convergence of FEM: the error should decrease as the number of degrees of freedom of the method increases.
We use this criterion to study the convergence of existing PINNs.}

\pranav{As promising future work, we plan to add a theoretical convergence analysis to our empirical convergence study of PINNs.
For classical PINNs consisting of an MLP accepting a coordinate input, we will follow existing analysis on the expressiveness of neural networks \cite{10.5555/3295222.3295371} to bound the approximation error of a PINN with regard to its width.
This expressiveness analysis will be combined with existing PINN convergence analysis that currently lacks degrees of freedom analysis (see \cref{sec:relatedwork}) to compute an error bound for the solution of a PDE that depends on the width and depth of the MLP in our PINN.
We will also combine this approach with well-known neural network scaling laws for PINNs to better understand the effects of other training parameters \cite{nnscalinglaws}.
While our current convergence analysis primarily focuses on PDE network width, with preliminary analysis on other hyperparameters such as PDE network depth and normal network, we are also excited to explore more sophisticated refinement strategies in future work.
}

\section{Acknowledgements}
This research is generously supported by a gift from Adobe Inc and a grant from the National Science Foundation (award \#2335493).

We acknowledge and thank the authors of the 3D assets
used in this paper. These models include Bob \cite{duck-mesh}, Bunny \cite{bunny-mesh}, Cat \cite{cat-mesh}, Dziuk \cite{Sugimoto2024}, Hand \cite{hand-mesh}, High-genus Sphere \cite{high-genus}, lion \cite{lion-mesh}, Moai \cite{moai-mesh}, Mushroom \cite{mushroom-mesh}, Plane \cite{plane-mesh}, Spot \cite{spot-mesh}, \cite{hammer-mesh}, Springer \cite{springer-mesh}.

\bibliography{ref}
\bibliographystyle{abbrvnat}

\appendix
\newpage
\appendix
\onecolumn
\clearpage
\section*{\centering Physics-Informed Neural Networks for PDEs on Curved Surfaces ---Appendices---}

\renewcommand{\contentsname}{}
\startcontents[sections]
\printcontents[sections]{ }{1}{\section*{\contentsname}\setlength{\cftsecnumwidth}{1.75em}}

\section{Further Experiments}
\subsection{Intricate Mesh}
\begin{figure}[!htb]
    \centering
    \includegraphics{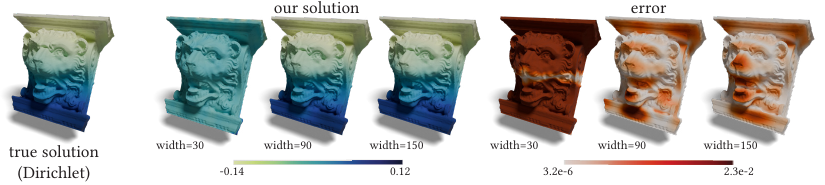}
    \caption{Our method can effectively solve PDEs even on mesh with intricate details.}
    \label{fig:intricate}
\end{figure}

In \cref{fig:intricate}, we show the capability of our method to solve the Helmholtz equation on a mesh with high curvature and fine details underscoring the effectiveness of our method.

\subsection{FEM Solution Dependence on Mesh Quality}
\begin{figure}[!htb]
    \centering
    \includegraphics{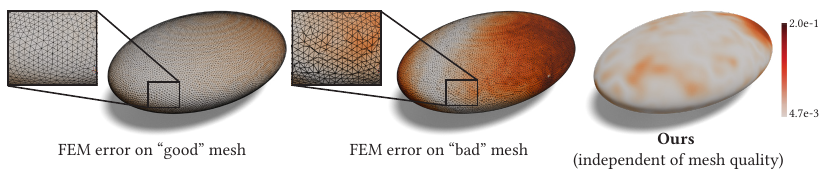}
    \caption{While the FEM error is dependent on the mesh quality -- error increases with mesh degradation, our method is independent of mesh quality.}
    \label{fig:bad_mesh}
\end{figure}

In \cref{fig:bad_mesh}, we solve the Poisson equation using FEM on meshes of varying quality and compare it with our PINN framework, which samples points directly from the exact surface. As mesh quality deteriorates, FEM accuracy degrades due to discretization errors, whereas our method remains robust, relying only on the ability to sample surface points rather than mesh quality.

\subsection{Further Ablations}
\begin{figure}[!htb]
    \centering
    \includegraphics{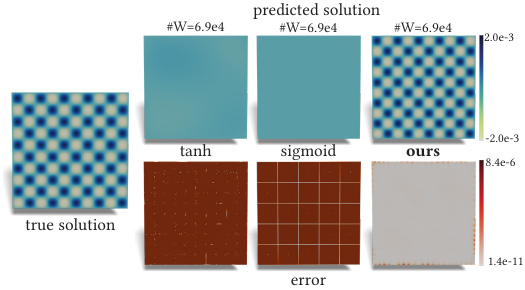}
    \caption{An ablation study where we compare the \emph{Siren} network, which uses $sine$ activation function with \emph{Tanh} and \emph{Sigmoid} activation functions on a 2D domain. The figure shows that \emph{Tanh} and \emph{Sigmoid} fail to solve for a high-frequency function.}
    \label{fig:activation_fn}
\end{figure}

In \cref{fig:activation_fn}, we show that the use of the \textit{Siren} network with \textit{sine} activation function results in a much lower error compared to using \textit{tanh} or \textit{sigmoid} activation functions. This experiment justifies our use of \textit{Siren} network both for the normal and the PDE network.

\begin{figure}[!htb]
    \centering
    \includegraphics{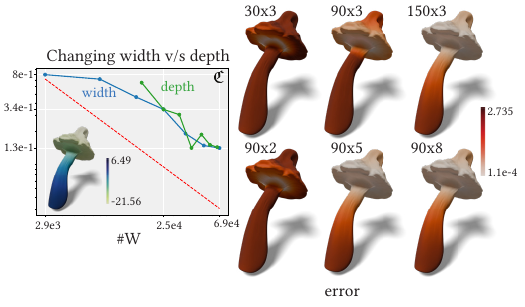}
    \caption{Result of an ablation study where we compare the final error first by changing the width, keeping the depth constant, and vice-versa. The figure shows that the final error in the solution to the PDE doesn't change when the network width is changed, compared to when the depth is changed.}
    \label{fig:width_depth}
\end{figure}

Since we use the number of tunable weights $\#W$ as our degrees of freedom, there are two ways to increase $\#W$ for an MLP -- by increasing the depth or by increasing the width of the MLP. In \cref{fig:width_depth}, we show that both choices result in the same error for equal number of tunable weights $\#W$. In all our experiments we choose to increase the width of the network to increase $\#W$.

\begin{figure}[!htb]
    \centering
    \includegraphics{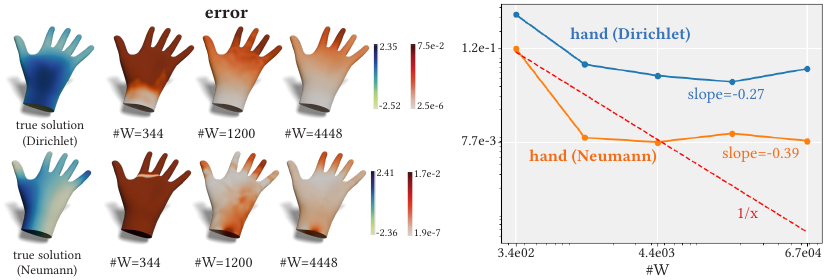}
    \caption{Figure showing how the error varies when the tunable parameters of the normal network are increased keeping the PDE network parameters constant.}
    \label{fig:normal_network_params}
\end{figure}

In \cref{fig:normal_network_params}, we show how the final error behaves as the normal network size is increased. We observe that the error first decreases and then stays constant showing that once the normal network is large enough to accurately learn the surface normals, the size of the normal network doesn't influence the final PDE error.

\subsection{Comparison with \cite{Sugimoto2024}}
\begin{figure}[!htb]
    \centering
    \includegraphics{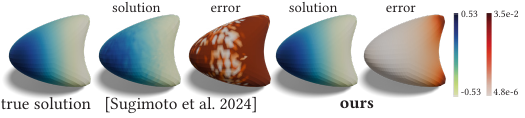}
    \caption{Comparison of our method with \cite{Sugimoto2024} which is a Monte-Carlo based method. We show that our method achieves a lower error when solving a screened Poisson equation on the Dziuk surface.}
    \label{fig:comparison_sugimoto}
\end{figure}

In \cref{fig:comparison_sugimoto}, we compare our results with those of \cite{Sugimoto2024}, a Monte Carlo–based approach. Our method demonstrates lower error when solving the screened Poisson equation on the Dziuk surface.

\subsection{Limitation Examples}
\begin{figure}[!htb]
    \centering
    \includegraphics{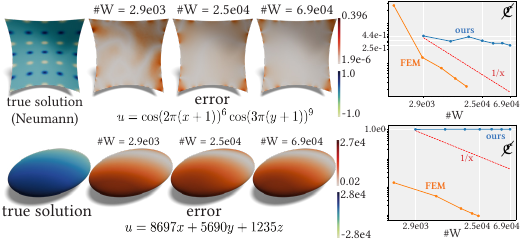}
    \caption{Figure showing the limitations of our work. Enforcing Neumann boundary is difficult at regions where the normal is discontinuous (for ex., at the corners of the heightfield). The method also fails if the solution $u$ has a huge variance.}
    \label{fig:limitations}
\end{figure}

\cref{fig:limitations} illustrates the limitations of our approach where our network fails to solve for the given PDE. Enforcing Neumann boundary conditions becomes challenging in regions where the normal is discontinuous (e.g., at the corners of the height field). Additionally, the method breaks down when the solution u exhibits large variance.

\section{Experimental Setup}
\label{sec:experiment_setup}
We set up experiments by first defining a pair of functions $u$ and $f$ on the given surface $\Omega$ that ensure homogeneous boundary conditions and the compatibility condition \eqref{eq:compatibility}.
We define analytic $u, f$ from a family of polynomial and trigonometric functions for surfaces whose normal field is known exactly.
For surfaces with boundary and no analytic form of the normal field, we use the eigenfunctions of the Laplace-Beltrami operator $\phi$ as the function $u$.
This gives
\begin{equation}
    \Delta_\Omega \phi_i = \lambda_i \phi_i = f,
\end{equation}
where $\phi_i$ and $\lambda_i$ are the $i^{th}$ eigenfunction and eigenvalue.\\
Similarly, for the Helmholtz equation
\begin{equation}
    \Delta_\Omega \phi_i + k^2 \phi_i = (\lambda_i + k^2)\phi_i = f.
\end{equation}
For surface meshes with no boundary and no analytic normal field, we define the source terms $f$ from a family of polynomial and trigonometric functions and compute $u$ using the cotangent Laplacian on a fine mesh. To ensure compatibility, we compute $\int_\Omega f dA$ using the mass matrix of the triangle mesh and subtract it from \(f\).

After defining the pairs $u$ and $f$, we set up our network with different widths and train each of them until complete saturation. Once every network is converged, we compute the relative $\ell_2$ error and plot it on a log-log plot to judge convergence as discussed in \cref{sec:convergence}. Note that for surfaces where the true analytical solution is not known, we compute $u_{\text{GT}}$ by solving
\begin{equation}
    -Lu_{\text{GT}} = Mf
\end{equation}
for the Poisson equation, where $L$ and $M$ are the cotangent Laplacian and the mass matrix of a fine triangular mesh, respectively. Similarly, for the  Helmholtz equation, we solve
\begin{equation}
    (-L + k^2I)u_{\text{GT}} = Mf.
\end{equation}
Given an analytical solution (or FEM solution with cotangent Laplacian) on a surface $\Omega$, we study the convergence patterns by training a PDE network $u_{\theta}$ of varying widths keeping the depth constant, and we plot the log of relative $\ell_2$ error with respect to the network parameters $\#W$ for each network after it's fully trained (\cref{fig:convergence_plots}).

\section{Computational Timings}
\label{sec:timings}
We ran all the experiments on an Intel i7 5.2GHz with a NVIDIA RTX 4080 GPU. We report the timings for each example shown in \cref{fig:error_vis} (Figure showing our solution and the error as the network width sizes are increased for Poisson and Helmholtz equations with
different boundary conditions) below. Note that we train each model for $150000$ iterations to ensure saturation but most models converge way before $150000$ iterations. Therefore, we report timing until convergence and also for full $150000$ iterations.

The computational time for solving the Poisson equation with Dirichlet boundary on the heightfield example (\cref{fig:error_vis}a) is as follows:
\begin{center}
\begin{tabular}{cccc}
\toprule
Width, Depth & Runtime (150000 iters) & Iterations to Convergence & Runtime to Convergence \\
\midrule
30, 3  & 26m 16s & 51039 & 8m 56s \\
90, 3  & 29m 22s & 69488 & 13m 42s \\
150, 3 & 37m 58s & 33319 & 8m 26s \\
\bottomrule
\end{tabular}
\end{center}
The computational time for solving the Helmholtz equation with Neumann boundary on the heightfield example (\cref{fig:error_vis}b) is as follows:
\begin{center}
    \begin{tabular}{cccc}
\toprule
Width, Depth & Runtime (150000 iters) & Iterations to Convergence & Runtime to Convergence \\
\midrule
30, 3  & 29m 49s & 61109 & 11m 48s \\
90, 3  & 35m 20s & 54123 & 12m 54s \\
150, 3 & 46m 24s & 41125 & 12m 40s \\
\bottomrule
\end{tabular}
\end{center}
The computational time for solving the Helmholtz equation with Dirichlet boundary on the hand example (\cref{fig:error_vis}c) is as follows:
\begin{center}
    \begin{tabular}{cccc}
\toprule
Width, Depth & Runtime (150000 iters) & Iterations to Convergence & Runtime to Convergence \\
\midrule
30, 3  & 19m 47s & 38630 & 5m 12s \\
90, 3  & 24m 50s & 33408 & 5m 46s \\
150, 3 & 32m 40s & 34051 & 7m 32s \\
\bottomrule
\end{tabular}
\end{center}

The computational time for solving the Helmholtz equation with Neumann boundary on the hand example (\cref{fig:error_vis}d) are as follows:
\begin{center}
    \begin{tabular}{cccc}
\toprule
Width, Depth & Runtime (150000 iters) & Iterations to Convergence & Runtime to Convergence \\
\midrule
30, 3  & 27m    & 35826 & 6m 46s \\
90, 3  & 34m    & 35995 & 8m 21s \\
150, 3 & 41m 46s & 26786 & 7m 38s \\
\bottomrule
\end{tabular}
\end{center}

The computational time for solving the Poisson equation on the bunny example (\cref{fig:error_vis}e) are as follows:
\begin{center}
    \begin{tabular}{cccc}
\toprule
Width, Depth & Runtime (150000 iters) & Iterations to Convergence & Runtime to Convergence \\
\midrule
30, 3  & 17m 23s & 46481 & 5m 29s \\
90, 3  & 22m 47s & 41163 & 6m 11s \\
150, 3 & 30m     & 29613 & 5m 53s \\
\bottomrule
\end{tabular}
\end{center}

The computational time for solving the Helmholtz equation on the spot example (\cref{fig:error_vis}f) are as follows:
\begin{center}
    \begin{tabular}{cccc}
\toprule
Width, Depth & Runtime (150000 iters) & Iterations to Convergence & Runtime to Convergence \\
\midrule
30, 3  & 18m 8s  & 48626 & 6m \\
90, 3  & 23m 10s & 29557 & 4m 39s \\
150, 3 & 29m     & 32337 & 6m 31s \\
\bottomrule
\end{tabular}
\end{center}

For normal and boundary normal networks, we train a fixed size network (width $64$ and depth $5$) for $1e7$ iterations for all examples. The normals are computed either analytically (if known) or using the mesh normals. The computational time for training the normal network is about $2h 40m$ ($1050$ iterations per second).

\subsection{Training Convergence Plots}
\begin{figure}[!htb]
    \centering
    \includegraphics{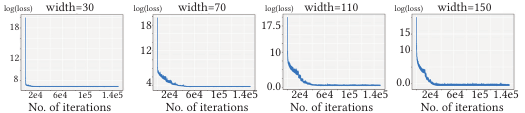}
    \caption{Figure showing the training loss for different $\#W$. The network usually converges around 50k iterations, but we train for 150k iterations to ensure complete saturation of the network.}
    \label{fig:loss_plots}
\end{figure}

\cref{fig:loss_plots} shows the training loss for different values of $\#W$ on the "bunny" model shown in \cref{fig:error_vis}e. The network typically converges around $50k$ iterations. However, we train for $150k$ iterations to ensure full saturation.

\section{Source Code Access}
Our complete code can be accessed here: \href{https://github.com/Pranav-Jain/NN_convergence.git}{code link}. The README.md contains information on setting up the environment and running scripts to reproduce the experimental results reported in the paper.

\end{document}